# Using Machine Learning to Identify Software Weaknesses from Software Requirement Specifications

Mounika Vanamala, Sean Loesch and Alexander Caravella

Department of Computer Science, University of Wisconsin-Eau Claire, Eau Claire, Wisconsin, USA
vanamalm@uwec.edu
seanloesch2002@gmail.com
alexander.caravella@gmail.com

## ABSTRACT

*Secure software engineering is crucial but can be time-consuming; therefore, methods that could expedite the identification of software weaknesses without reducing the process' efficacy would benefit the software engineering industry and thus benefit modern life. This research focuses on finding an efficient machine learning algorithm to identify software weaknesses from requirement specifications. The research uses the CWE repository and PROMISE_exp dataset for training. Keywords extracted using latent semantic analysis help map the CWE categories to PROMISE_exp. Naïve Bayes, support vector machine (SVM), decision trees, neural network, and convolutional neural network (CNN) algorithms were tested, with SVM and neural network producing reliable results. The research's unique contribution lies in the mapping technique and algorithm selection. It serves as a valuable reference for the secure software engineering community seeking to expedite the development lifecycle without compromising efficacy. Future work involves testing more algorithms, optimizing existing ones, and improving the training sets' accuracy.*

## 1. INTRODUCTION

Secure software engineering is essential to contemporary infrastructure. Most of the luxuries and efficiency of the modern world can be attributed to the safe construction and operation of software and code. Society, as we know it today, would fail without software. There are over twenty billion IoT (Internet of Things) devices (devices with the ability to transmit data over the Internet) as of 2019 [1]. Such a precedent requires the software that is utilized by the masses to be secure, and a response was needed. Thus, the Secure Development Lifecycle was born.

Secure Development Lifecycle (SDL) is the process in the Software Development Lifecycle (SDLC) in which security artifacts are prioritized first [2]. This model is paramount in building safe and secure software; however, as with any precautionary measures, security is purchased at the price of time and labor. Such an observation is as old as software development itself – even a former member of tech giant Microsoft's central security team acknowledged that "reviewing thousands of files can be slow and tedious" [3].

One of the first steps of both the Secure Development Lifecycle and the Software Development Lifecycle is a software requirements specification (SRS). Put in simple terms, an SRS is an enumeration of functional and nonfunctional requirements by the client. Functional requirements typically address functional features, and nonfunctional requirements could address things such as appearance and ease of use. Each functional and nonfunctional requirement typically comes with a description and its subcategory. The PROMISE_exp repository includes a large preponderance of such software requirements from a myriad of verified sources [4] and will be used in this work.

An SRS is a microcosm of the Software Development Lifecycle in terms of its tediousness — it requires much deliberation and negotiation between the parties involved in the creation and utilization of software, typically between the client and the development company or team. This process is called the "requirements analysis." In the Secure Development Lifecycle, risk assessment occurs contemporaneously with requirements analysis to determine what potential hazards could occur during development. This process is not perfect, however. It is reasonable to assume that mistakes or information left out in the requirements elicitation process could lead to weaknesses in the software if they pass unnoticed. A tool to aid in discovering potential future weaknesses so that developers can approach building software more cautiously would save time and effort and lead to safer and more secure software. Thus, our goal is to supplement the SDL using a machine learning algorithm that will help to predetermine software weaknesses from requirements elicitation documents.

The Common Weakness Enumeration (CWE) is a categorical system listing common software weaknesses and vulnerabilities developed by the software development community and maintained by the MITRE corporation [5]. It provides an intuitive and common language to categorize and succinctly describe distinct software weaknesses according to their definition and risk. This allows software developers to quickly identify and develop remedies for erroneous software.

The CWE repository can be mapped (related) to the PROMISE_exp repository through text classification. Text classification is a form of supervised machine learning that allows for the categorization of text into organized groups. According to Manning et al., "this type of learning is called supervised learning because a supervisor (the human who defines the classes and labels training documents) serves as a teacher directing the learning process" [6]. The application of a text classification algorithm on both the CWE and PROMISE_exp repositories allows for the relation between the two to form, and thus opens the door for various other machine learning algorithms to predict software weaknesses from SRSs.

In this work, we will scrutinize how to optimize the classification of software requirements and thus produce a practical algorithm that will allow for the accurate prediction of software weaknesses from requirement elicitation documents. Forming such an algorithm begins with what is called Latent Semantic Analysis. Latent Semantic Analysis (LSA) is a method for bridging the gap between the syntactic "structure" of language and the semantic "meaning" of the same language [7]. It is an intricate process; however, the simplest way to understand its function is to examine the Bag of Words model that utilized in its preprocessing.

To summarize, a vectorization technique should be used to represent keywords numerically so they may be processed. The vectorization technique used for our LSA construction was the Bag of Words (BoW) model. The BoW is one of the most popular methods for object categorization representation and involves selecting key objects (e.g., words) from a larger parent object (e.g., a line in a text file) in a manner that allows the larger object to be sufficiently represented by the abstracted key objects [8]. As mentioned previously, Latent Semantic Analysis can be applied to text files. Thus, applying an LSA algorithm to the descriptions of weaknesses in the CWE repository and the lines in the PROMISE_exp dataset allows them to be mapped to one another with a high degree of accuracy. Once that is completed, the IDs of each CWE weakness (or each "CWE ID") can be further abstracted to be the encapsulating categories under which each CWE ID falls. From there, each CWE category can be mapped to each line in the PROMISE_exp dataset and the resulting file can be used as a training set for machine learning algorithms. Such machine learning algorithms (described in "Methodology") would use the training set to accurately predict and identify categories of software weaknesses from software requirement specifications.

## 2. LITERATURE REVIEW

Based on our research in identifying proper machine learning techniques which are effective in determining correlations between the Common Vulnerable Exposures repository (CVE) and software requirements we branched our work based on the articles "Recommending Attack Patterns for Software Requirements Document," "Analyzing CVE Database using Unsupervised Topic Modelling," and "Topic Modeling and Classification of CVE Databases."

Prior to this research, large databases which held information about a variety of attack patterns were developed to help software developers identify weaknesses within their code. As explained in "Recommending Attack Patterns for Software Requirements Document" a database known as CAPEC was developed from CVE data to provide more assistance to software developers while they were creating the software requirements for their programs [9]. To achieve this goal, the authors used the LDA algorithm for topic modeling to test how accurate their ideas of topic modeling were in identifying keywords between CAPEC and SRS documents. While going through this process they also used a metric process called Cosine Similarity which bases its math on how similar documents are while not taking their sizes into account and then creates two vectors based on each document and projects it onto a multi-dimensional space [10]. However, this methodology is stated to be heavily dependent on the text describing the system due to it looking for specific keywords. Once the algorithm was run for topic modeling it was found to be a success in identifying relationships between CAPEC attacks as well as patterns from SRS documents.

Along with helping to develop CAPEC, the CVE repository was also further studied to analyze vulnerabilities reported in it based on the article "Analyzing CVE Database using Unsupervised Topic Modelling" [11]. From the document, it focuses on using topic modeling and the LDA algorithm to discover topics within the CVE database and mapped them to the OWASP top 10 risks. To map these topics using LDA they went through multiple phases being broken down into data preprocessing, text cleaning, bigram, and trigram modeling, building the topic model, visualizing the topics and associated keywords, and finally mapping topics to the OWASP-Top 10. Once this process was completed, they were able to identify growing trends in cyber-attacks based on taking data from 5-year intervals starting from 1999 and going to 2019. Once these groupings were compared it was found that there was a large similarity between the OWASP top-10 recommendations, and the growing vulnerabilities reported from CVE. Based on the document's results, they found it successful due to the striking similarities between the OWASP top-10 and the growing reports on the CVE database.

In a follow up to the article about analyzing the CVE database there was another used which referred to "Topic Modeling and Classification of CVE Databases" [12]. From this research, by implementing a LDA machine learning algorithm they further identified correlating topics between information from the CVE database to the OWASP Top-10. By using statistical analysis to determine the results of both lists, they identified remarkably similar values in standard deviation and coefficient of variance based on when they originally manually mapped the results compared to when the used automated mapping for this project. From the results, it proved that it was possible to replicate data that was previously manually mapped to being automatically mapped by machine learning. By using this technique of automatic mapping, they are hoping to develop a framework to automate the process of mapping the vulnerability reports and set a security standard based on its results.

Going along with this prior research, connections between CVE and CAPEC were also conducted to link both databases automatically using Artificial Intelligence. By using this technique, they could link information from the database together and better understand what topics correlated to what attacks. Algorithms such as SBERT, TF-IDF, and USE linked these databases based on CAPEC APEC ids and CVE ids. This topic in the research coincides closely with what we plan to do with ours due to wanting to link software requirement ids with CVE ids which connect security information based on the topic of the software requirement.

Another research which also focused on a model to quantify software with analysis alerts and software metrics worked by Siavvas et al [13]. From the research, the authors worked on using systems like CWE along with a multifaceted model to evaluate security. Based on tests they ran on java applications they were able to classify levels of security as well as a list for developers to work on. Thanks to this information, it saved time and mitigated some process time needed to work with the code. Much like with this work, we are hoping to create a list of what the specific security problems are instead of the levels of security cyberattacks could affect.

Based on our research being conducted with CWE we decided to research machine learning techniques and found an article describing the study of four machine learning techniques for classification written by Amr E. Mohamed [14]. Based on the work he did, his focus was on finding the key ideas of each technique and including their advantages and disadvantages in working with them. From the study, he chose to work with support vector machines (SVM), artificial neural networks, decision trees, and K-nearest-neighbors. Starting with SVM, he found that some of its major advantages were its high scaling functions as well as the model complexity and errors being easy to control. Although one disadvantage they found was its difficulty in interpreting information based on how it is read, a result that has been observed in other research [15]. With neural networks, he found that their ability to learn made them immensely powerful and flexible to use and has been useful in solving classification, clustering, and regression problems. Some weaknesses though included that the successes depended on the quantity of the data and that there is a lack of clear guidelines due to it being based on trial and error. With decision trees, he found that they can easily provide information that can be very comprehensible for readers however, they can easily overfit the data and are not as useful for tasks involving regression. Finally, with K-nearest-neighbors, he found that the speed of the algorithm is amazingly fast and easy to implement. However as other research corroborates, it is sensitive to irrelevant features and can be computationally expensive [16]. The information in the paper was particularly useful in our research due to providing information on what algorithms would be best when running these machine learning processes after we used topic modeling on our code.

The research focused on vectorization of natural language processing written by Krzeszewska et al [17] was studied to understand and provide linguistic context when processing natural language to classification in machine learning. By using a continuous Bag of Words and skip-gram, they obtained classification and keywords based on the topic discussed in the text document they used. used. From here, they were able to take the information and create a summary statistic based on categories they wanted to separate these topics into by using KNN and Naïve-Bayes to classify the information. Once they compiled the data, they found it was considered reliable. They found that even with a large dataset there was no negative impact on the vectorization methods used to find these conclusive results. Based on this information, they found that there was a good relationship between the vectorization methods and the classification performance. While we conducted our own research, we used techniques from the article to find our results by matching IDs between CWE and our dataset of software requirements with help from the Bag of Words vectorization model.

## 3. METHODOLOGY

In this phase of our research, our goal was to test different machine learning models to identify which algorithms would most accurately predict which software requirement would be appropriately mapped to the correct cybersecurity vulnerability based on the CWE database.

While conducting this research, we created machine learning models using Naïve Bayes, support vector machine (SVM), decision trees, a neural network, and a convolutional neural network (CNN).

To form the algorithm, we first had to create the training set that said algorithms would use. To do this, we began by performing Latent Semantic Analysis (LSA) to form the algorithm. In LSA, we bridge the gap between the syntactic structure and semantic meaning of language. To simplify the understanding of LSA, let's consider the preprocessing step that involved the Bag of Words model.

To summarize, we utilized a vectorization technique to represent keywords numerically for processing. The vectorization technique employed for constructing our LSA was the Bag of Words (BoW) model. The BoW model is commonly used for object categorization representation, where key objects (e.g., words) are selected from a larger parent object (e.g., a line in a text file) to effectively represent the larger object.

Having mentioned the applicability of Latent Semantic Analysis to text files, we applied the LSA algorithm to the descriptions of weaknesses in the CWE repository and the lines in the PROMISE_exp dataset. This process allowed for a highly accurate mapping between them. Subsequently, the CWE weaknesses were further abstracted into CWE categories, each represented by a CWE ID. These CWE categories were then mapped to each line in the PROMISE_exp dataset. The resulting file served as a training set for machine learning algorithms described in the "Methodology" section. Once we had a complete CSV file which included the software requirement, the CWE vulnerability description, and the CWE vulnerability ID we were then ready to use the machine learning algorithms to find how accurate our results were. These algorithms utilized the training set to predict and identify software weakness categories from software requirement specifications. When testing and training the algorithms we used three different sets. The first was an 80-20 training/test split, then we used a 70-30 split, and then we used a 60-40 split.

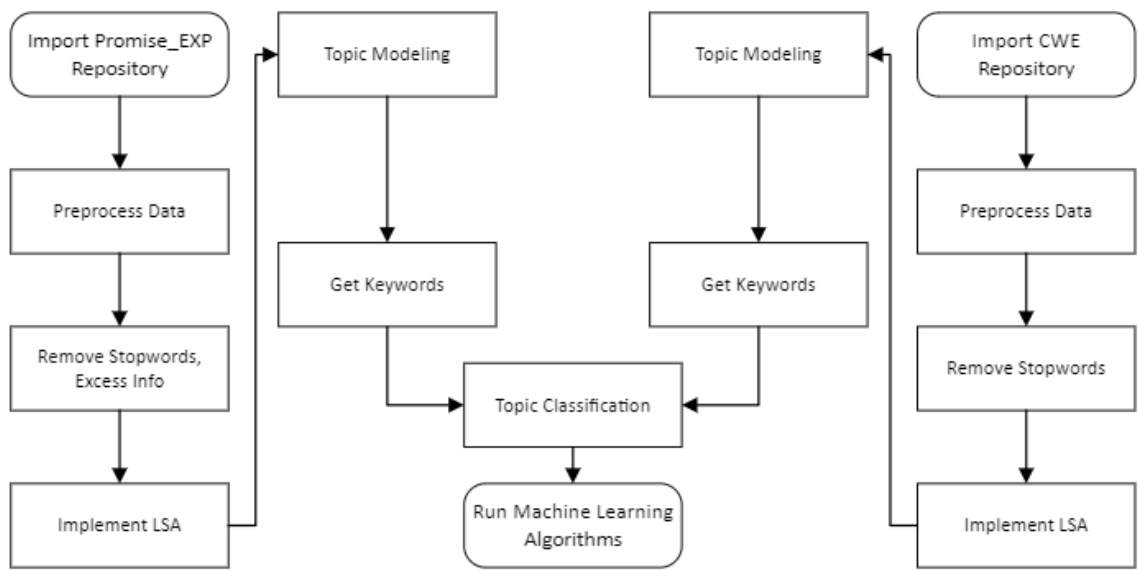

Figure 1. Flowchart of methodology.

## 4. RESULTS AND DISCUSSION

In this section, we will review the findings from the testing of each algorithm and provide reasoning regarding why the ones that did not work did not. As mentioned previously, the algorithms that were tested in total were: Naïve Bayes, SVM, decision trees, neural network, and convolutional neural network (CNN). Among these algorithms, the ones that provided reliable results were the SVM and the neural network.

The SVM algorithm yielded an average accuracy of ≈0.642. To be precise, for the 80-20, 70-30, and 60-40 training/test splits, accuracies of ≈0.648, ≈0.643, and ≈0.636 were obtained, respectively. This moderate but not exceptionally high accuracy rate makes the SVM algorithm a clear candidate for use – however, additional modifications would be needed to make it viable to compete with the neural network [18]. Over dozens of repetitions, the neural network never yielded an accuracy below 0.86, with most runs ending with an accuracy higher than 0.90 and achieving a high of ≈0.992. The neural network had a batch size of 32 samples and went through 10 epochs before reaching its final accuracy for any run. This means the neural network is currently the optimal algorithm to use.

Although the success of the results garnered from the algorithms that did work was significant, it is important to make note of the machine learning algorithms that did not work with the CWE/PROMISE_exp mapped training dataset. The algorithms that did not work were Naïve Bayes and decision trees. The convolutional neural network (CNN) hasn't been developed enough to be able to conclude anything regarding its reliability currently.

Naïve Bayes can be split into three variants: Gaussian Naïve Bayes, Multinomial Naïve Bayes, and Bernoulli Naïve Bayes. Only the Gaussian Naïve Bayes and Multinomial Naïve Bayes were used since the Bernoulli variant deals with data with binary features (i.e., each feature can only take on two values) [19], [20]. The attempt to use the Multinomial variant resulted in an error due to the dataset not being able to assume the correct format for the variant to function properly. Multinomial Naïve Bayes expects data in the form of counts, where each feature represents the frequency of occurrence of a particular term or word in a document or sample [21]. The type of data collection used for creating the training set could not include allocating data in this way, rendering the Multinomial variant unusable as well. As for the Gaussian variant, results showed a low average accuracy (≈0.12) over the results from 80-20, 70-30, and 60-40 data splits. This was most likely because the Gaussian variant works best with normally distributed data [22]. The features of the training dataset were not at all normally distributed, which was the culprit for its bad results. Thus, since the Gaussian and Multinomial variants were thrown out and the Bernoulli variant was not applicable, Naïve Bayes was thrown out in its entirety.

The way the dataset was structured was the reason the decision trees algorithm was discarded as well. Decision Trees rely on identifying patterns and relationships in the data to make accurate predictions [23]. Since the value of the target variable in every row was unique, there were no repeating patterns within the dataset for the algorithm to discern, yielding a low accuracy (≈0.10).

## 5. CONCLUSION AND FUTURE WORK

In this work, we used the results from applying Latent Semantic Analysis to the content stored in the Common Weakness Enumeration and the PROMISE_exp repositories to map the best fitting CWE weakness category to each line of the PROMISE_exp repository. The result of this mapping became the dataset that would be used in our experiments. This training set was created to allow us to build an algorithm that could predict the potential development of software weaknesses. Upon its completion, this algorithm could be fed software requirements specifications and accurately predict which weaknesses could occur in the future so that secure software developers could expedite creating software in a secure manner which the same level of efficacy.

We conducted an experiment to compare the performance (accuracy/precision) of various supervised machine learning algorithms to arrive at the most optimal algorithm for the task. We found that a neural network with 10 epochs and a batch size of 32 showed the best results in precision, troughing at no less than an accuracy value of 0.86 and peaking at 0.992, while remaining above 0.9 on average.

Secure Software Engineering is an onerous process. Identification of potential software weaknesses from requirements elicitations using an accurate machine learning algorithm would save developers a significant amount of labor and time in finding weaknesses in their code, and thus allow software of the same quality as before the existence of such an algorithm to be produced at an expedited rate. We hope this study will help developers who want to optimize the secure development of software and to produce high-quality work at a faster rate. The training dataset developed could also serve as a reference for other studies, allowing other researchers to draw their own unique and useful conclusions regarding the optimal algorithm for use.

For future work, we intend to optimize currently working algorithms to provide new results with higher accuracy rates, test new algorithms in the search to find the optimal algorithm to use for this task and correct the training set used in the algorithms for error and optimize it.